\documentclass[pra,amsmath,amssymb,twocolumn,superscriptaddress,showpacs]{revtex4-2}

\usepackage{graphicx}
\usepackage{dcolumn}
\usepackage{bm}
\usepackage{hyperref}

\begin{document}
	
	\preprint{APS/123-QED}
	
	\title{Higher-order topological phases in bilayer phononic crystals and topological bound states in the continuum}
	
	\author{Xiao-Yu Liu}
	\affiliation{School of Physical Science and Technology, \& Collaborative Innovation Center of Suzhou Nano Science and Technology, Soochow University, Suzhou 215006, China}
	
	\author{Yang Liu}%
	\affiliation{School of Physical Science and Technology, \& Collaborative Innovation Center of Suzhou Nano Science and Technology, Soochow University, Suzhou 215006, China}
	
	\author{Zhan Xiong}
	\email{xiongzhan@zjnu.edu.cn}
	\affiliation{College of Physics and Electronic Information Engineering, Zhejiang Normal University, Jinhua 321004, China}
	\affiliation{Key Laboratory of Optical Information Detecting and Display Technology, Zhejiang Normal University, Jinhua 321004, China}
	
	\author{Hai-Xiao Wang}%
	\email{wanghaixiao@nbu.edu.cn}
	\affiliation{School of Physical Science and Technology, Ningbo University, Ningbo 315211, China}%
	
	\author{Jian-Hua Jiang}%
	\email{jianhuajiang@suda.edu.cn}
	\affiliation{School of Physical Science and Technology, \& Collaborative Innovation Center of Suzhou Nano Science and Technology, Soochow University, Suzhou 215006, China}%
	\affiliation{Suzhou Institute for Advanced Research, University of Science and Technology of China, Suzhou 215123, China}%
	
	\date{\today}
	
	\begin{abstract}
		Recent studies on the interplay between band topology and layer degree of freedom provide an effective way to realize exotic topological phases. Here we systematically study the $C_6$- and $C_3$-symmetric higher-order topological phases in bilayer spinless tight-binding lattice models. For concreteness, we consider bilayer phononic crystals as the realizations of these models. We find that for mirror-symmetric-stacking bilayer lattices, the interlayer couplings control the emergence and disappearance of the topological bound states in the continuum where we consider the corner states as possible bound states in the bulk continuum. For the bilayer phononic crystals formed by two different lattices with identical symmetry, the band topology is determined by both the band topology of each layer as well as their mutual couplings. The bilayer phononic crystals experience a phase transition from nontrivial to trivial band topology when the interlayer couplings are gradually increased. Our work unveils the rich physics and topological phases emerging in bilayer lattice systems that can be used to engineer interesting phenomena and induce emergent topological phases.
	\end{abstract}
	
	\maketitle
	
	\section{\label{sec:level1}Introduction}
	Exploring topological phases in phononic systems gives birth to the  field of topological phononics~\cite{toporev1,toporev2,toporev3,toporev4,toporev5,toporev6}. Benefiting from flexibility and scalability in the design and fabrication, phononic crystals provide a powerful and versatile platform to realize various topological phases, which in turn largely enrich the manipulation of the acoustic and elastic waves. Recently, the discovery of higher-order topological phases has largely expanded the classification of the topological phases of matters~\cite{hotrev}. In contrast to the conventional topology, higher-order topology manifests itself in multidimensional topological boundary states (such as edge and corner states) beyond the conventional bulk-boundary correspondence. Various higher-order topological phases, including quadruple topological insulators~\cite{science.357.61benalcazar,nature555.342Serra-Garcia,nat.commun.11.65zhang,phys.rev.lett.124.206601qi}, octupole topological insulators~\cite{nat.commun.11.2108ni,nat.commun.11.2442xue}, Wannier-type higher-order topological insulators~\cite{phys.rev.lett.120.026801ezawa,nat.mater.18.108xue,nat.mater.18.113ni,phys.rev.b.98.045125ezawa,phys.rev.lett.126.156401zhang,phys.rev.lett.122.244301xue,adv.mater.31.1904682zhang,appl.phys.lett.117.113501yang}, higher-order Weyl and Dirac semimetals~\cite{higherWeyllayer1,higherWeyllayer2,higherWeyllayer3,higherWeyllayer4,higherWeyllayer5,higherWeyllayer}, have been theoretically proposed and experimentally demonstrated to host wave localization in multidimensional boundaries. For example, two-dimensional higher-order acoustic topological insulators with both gapped edge states and in-gap corner states offer an unprecedented way to realize wave localization at the edges and corners in a dimensional hierarchy manner, which may find potential applications on topological routing of acoustic waves. 
	
	On the other hand, there has been growing interest in introducing the layer degree of freedom to topological phononics, which likely give rise to unconventional topological phases. The layer degree of freedom with the tunable interlayer couplings provides an efficient tool to design phononic states. For example, the layer-stacking approach offers a scheme to enrich the spatial symmetry, which plays a key role in the design of three-dimensional acoustic topological insulators with Dirac hierarchy~\cite{bilayerhierarchy1,bilayerhierarchy2,bilayerhierarchy3,bilayerhierarchy4,bilayerhierarchy5}, and higher-order Weyl and Dirac semimetals~\cite{higherWeyllayer1,higherWeyllayer2,higherWeyllayer3,higherWeyllayer4,higherWeyllayer5,higherWeyllayer}. Furthermore, the combination of the layer and valley degrees of freedom gives rise to a rich topological phase diagram, as shown in honeycomb phononic crystals~\cite{valleybilayer1,valleybilayer2,valleybilayer3,valleybilayer4,valleybilayer5}. In Lieb lattice phononic crystals, an acoustic spin-Chern insulator is proposed by introducing a layer degree of freedom with the proper interlayer coupling~\cite{spinbilayer1,spinbilayer2}, in which the layer plays a role of the pseudospin degree of freedom. Very recently, it was reported that layer-stacked structures with mirror symmetry provides a useful approach to turn the boundary states of any topological monolayer model into topological bound states in the continuum~\cite{bilayerBIC,higherWeyllayer6}. To date, the interplay between the layer degree of freedom and the higher-order topological phases, however, remains largely unexplored, especially for bilayer structures composed of two distinct monolayers.    
	
\begin{figure}[htbp]
		\centering\includegraphics[width=\columnwidth]{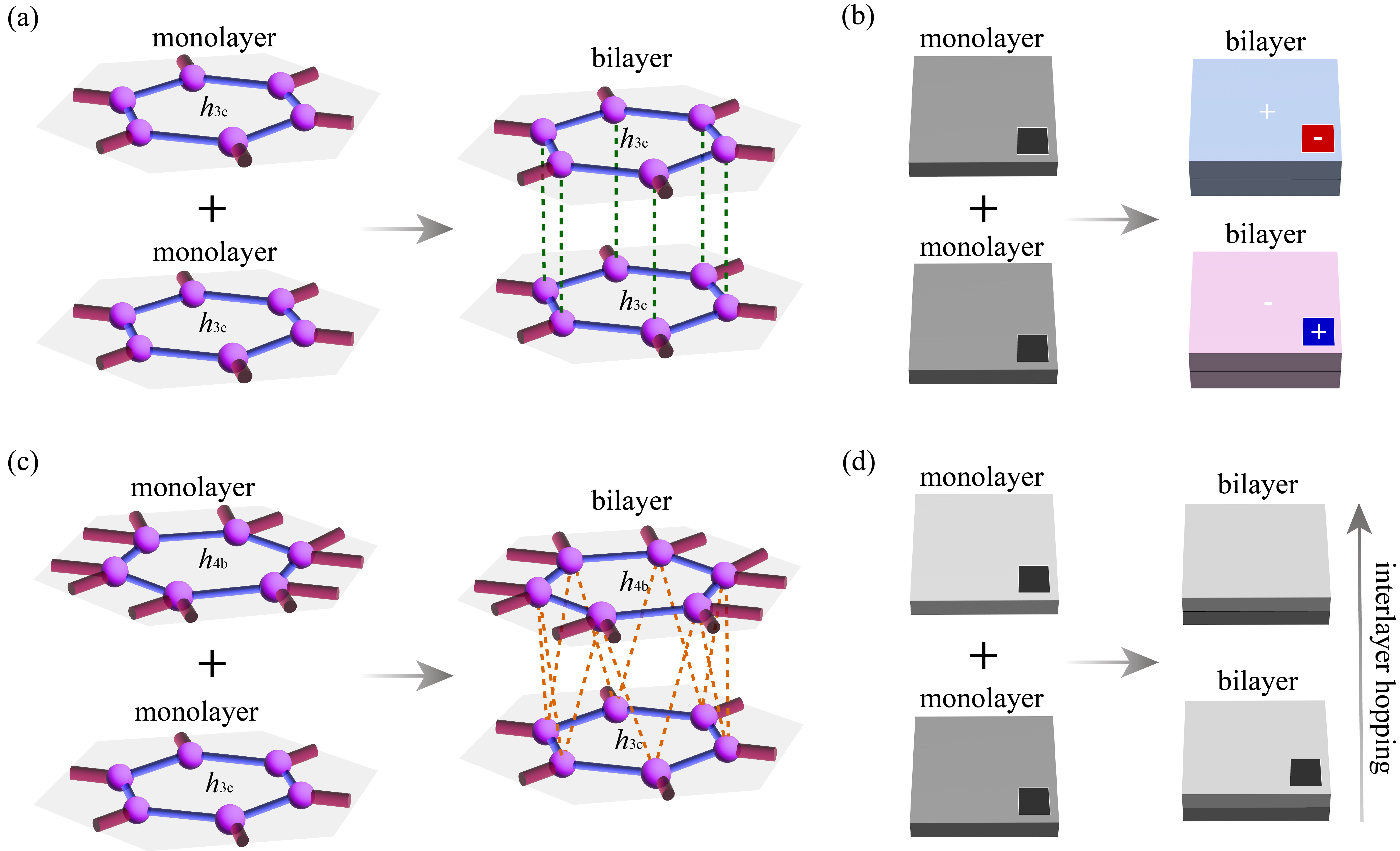}
		\caption{(a) Bilayer phononic crystals (with mirror symmetry) formed by stacking two identical monolayers $h_{3c}$. (b) Mirror-stacked bilayer phononic crystals supporting two different topological bound states in the continuum. (c) Bilayer phononic crystals (without mirror symmetry) formed by stacking two distinct monolayer $h_{4b}$ and $h_{3c}$. (d) Stacked bilayer phononic crystal formed by monolayers $h_{4b}$ and $h_{3c}$, of which the phase transition is triggered by tuning interlayer couplings.}
		\label{Fig_0}
	\end{figure}

	To fill this gap, we systematically study the bilayer phononic crystals consisting of either identical or distinct monolayers with higher-order band topology. For instance, taking the monolayer with $C_6$ symmetry into consideration, there are two categories of $C_6$-symmetric higher-order topological insulators, labeled as $h_{4b}$ and $h_{3c}$, which are classified according to the Wannier centers of the bands below the gap~\cite{topocharater1}. When two identical monolayers, e.g., $h_{3c}$ and $h_{3c}$, are stacked together, a bilayer structure with mirror symmetry forms [see Fig.~\ref{Fig_0}(a)]. Remarkably, the presence of mirror symmetry enables a classification of the states into two subspaces according to their mirror eigenvalues, i.e., even or odd. By tuning the interlayer couplings, the topological boundary states in the even subspace can be tuned into the bulk continuum of the odd subspace, giving rise to the topological bound modes in continuum [see sketch in Fig.~\ref{Fig_0}(b)]. On the other hand, by stacking two distinct monolayers, e.g., $h_3c$ and $h_{4b}$, bilayer structure without mirror symmetry can be formed [see Fig.~\ref{Fig_0}(c)]. In this case, by tuning the interlayer coupling, the band structure of the bilayer phononic crystal undergoes the process of band hybridization and reorganization, leading to unique topological transitions and emergent higher-order topological phases [see the sketched in Fig.~\ref{Fig_0}(d)]. 
	
	For convenience, we divide the bilayer phononic crystals into two types, according to the symmetry of layer stacking. The first type is the mirror-stacked bilayer phononic crystals (MBPCs) that consist of two identical monolayer phononic crystals. The second type is the heterogeneous-stacked bilayer phononic crystals (HBPCs) that consist of two monolayer phononic crystals with distinct structure and band topology. In the following sections, we will elaborate on the properties of these two types of phononic crystals. We pay special attention to the role of the interlayer couplings, the symmetry and band topology, as well as the topological boundary states. Specifically, in Sec.~II we study the mirror-stacked bilayer phononic crystals and the topological boundary states in the bulk continuum. In Sec.~III we discuss the heterogeneous-stacked bilayer phononic crystals and their band topology as well as the topological corner states. We conclude and discuss all the results as a whole in Sec.~IV. In all calculations, we use the commercial finite-element software COMSOL Multiphysics to obtain the phononic dispersions and eigenstates.
	
	\section{Mirror-stacked bilayer phononic crystals}
	
	In this section, we focus our attention on the MBPCs and its physical results. Throughout this work, the phononic crystals are made of cavity-tube structures, in which the height and diameter of the individual cavity is $h_1=38$mm and $D=16$mm, respectively. Physically, the cavity resonators mimic atomic orbitals and the narrow tubes introduce hoppings between them. Hence, cavity-tube structure-based phononic crystals provide an ideal platform to mimic the corresponding tight-binding models. 
	
	\subsection{$C_6$-symmetric MBPC formed by Wu-Hu's lattice ($h^{(6)}_{3c}$ and $h^{(6)}_{3c}$)}
	
\begin{figure}[htbp]
		\centering\includegraphics[width=\columnwidth]{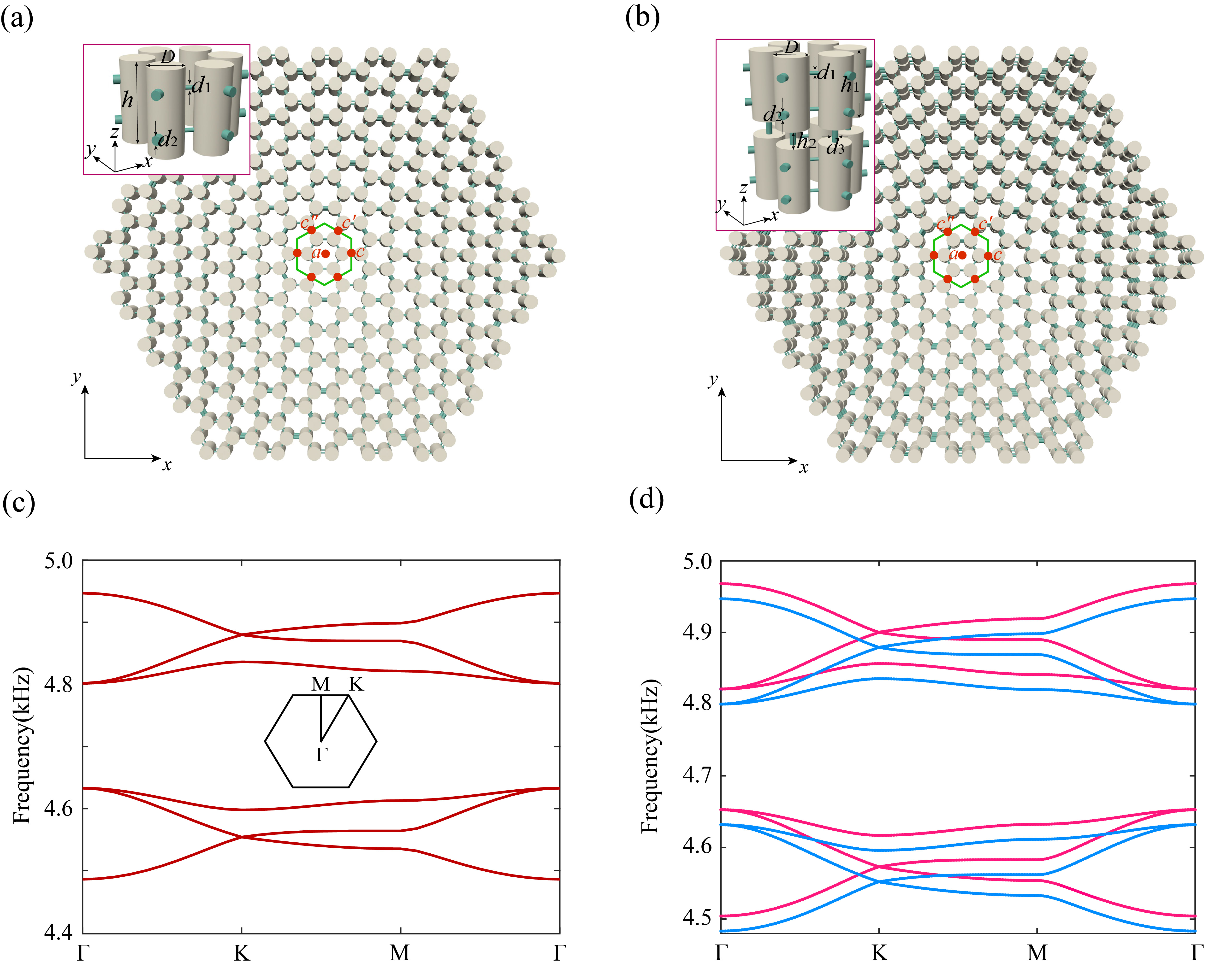}
		\caption{(a) Schematic of monolayer phononic crystals with $h^{(6)}_{3c}$ configuration. Inset: the side view of the primitive cell. (b) MBPCs formed by two identical monolayers with $h^{(6)}_{3c}$ configuration. (c) Band structure of monolayer phononic crystals. (d) Band structure of the MBPC formed by two identical monolayers with $h^{(6)}_{3c}$ configuration.}
		\label{Fig_1}
	\end{figure}

	As a starting point, we utilize a two-dimensional six-fold ($C_6$) rotation symmetric phononic crystals as a monolayer to work on. As shown in Fig.~\ref{Fig_1}(a), the lattice constant is $a=62$mm. The intra-cell and inter-cell couplings are realized by the air tubes with a diameter $d_1=2.0$mm and $d_2=4.2$mm, respectively. Note that such a $C_6$-symmetric lattice with nontrivial band topology was first proposed in Ref.~\cite{Wu-Hu-model} (denoted as Wu-Hu's lattice), and has been later on extensively studied in the literature. Next, a MBPC forms by stacking such a monolayer of Wu-Hu lattice along $z$-direction with distance $h_2=9.5$mm,  [Fig.~\ref{Fig_1}(b)]. By tuning the diameter $d_3$ of the connecting air tube, the interlayer coupling can be finely controlled. 
	
	For the monolayer phononic crystals, it is seen that there are a total of six acoustic bands, and a complete band gap separates them into two groups [Fig.~\ref{Fig_1}(c)]. Since the intercell coupling is larger than the intracell coupling ($d_2>d_1$), we remark that such a complete band gap is of a nontrivial topological nature. In the view of the Wannier configuration, the Wannier centers of the lowest three bands are located at positions $c$, $c'$, and $c''$, respectively [also see the green hexagon in Fig.~\ref{Fig_1}(a)]. For convenience, we use the notation $h^{(n)}_{mW}$ to characterize the monolayer phononic crystals with $C_n$ symmetric with $m$ filled bands and has Wannier centers at the maximal Wyckoff position $W$. Hence, the monolayer phononic crystals in Figs.~\ref{Fig_1}(a) and~\ref{Fig_1}(c) are termed as $h^{(6)}_{3c}$. For the mirror-stacked bilayer phononic crystals with $h^{(6)}_{3c}$ [Fig.~\ref{Fig_1}(d)], it is seen that there are twelve bands, which according to their mirror parities, can be divided into two sets. The blue (red) bandset refers to the bands with even (odd) mirror parity. Obviously, there exists an energy offset between two bandsets due to the interlayer coupling, and the band structure of each bandset is identical to that of the monolayer phononic crystal.  
	
\begin{figure*}[htbp]
		\centering\includegraphics[width=6.8in]{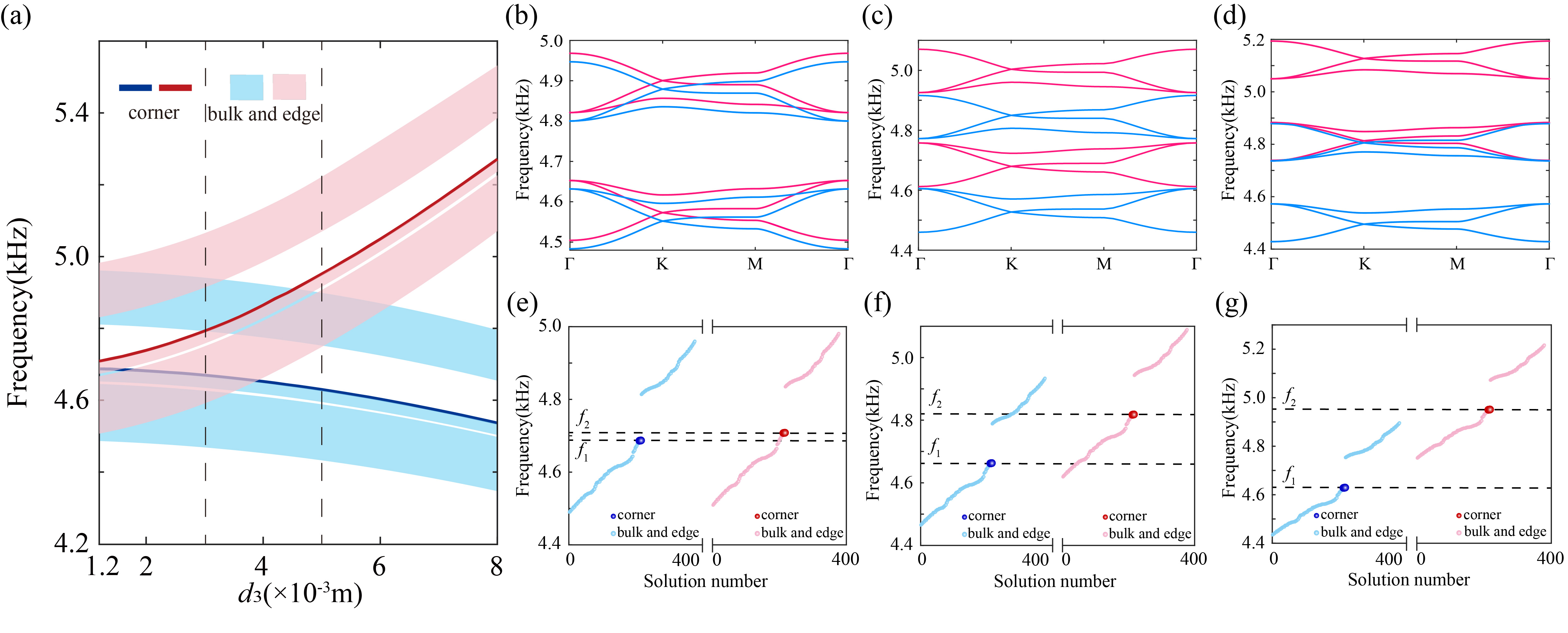}
		\caption{(a) The eigen spectra of MBPC versus the diameter $d_3$ of the connecting air tubes. (b-d) The band structure of the MBPCs with diameter of the connecting air tubes (b) $d_3=1.2$mm, (c) $d_3=3.0$mm, (d) $d_3=5.0$mm. (e-g) The corresponding of eigen spectra of the MBPC in (b-d), where the blue and red points refer to the acoustic states with even and odd mirror parities, respectively.}
		\label{Fig_2}
	\end{figure*}

	The MBPCs inherit the original band topology of the monolayer, which, in this case, is a higher-order topological phase. It is predicted that the corner states can move continuously into and out of the two-dimensional bulk continuum of opposite parity by tuning the interlayer couplings, which leads to the appearance and disappearance of the topological bound states in the continuum. To this end, we plot the energy spectra of the finite-sized MBPC versus the diameter $d_3$ of the connecting air tube in Fig.~\ref{Fig_2}(a). The shadow light blue (red) area and the blue (red) line refer to the bulk and corner states of MBPC with even (odd) parity, respectively. For each subspace, it is clearly seen that a corner state emerges as the manifestation of the higher-order band topology. Remarkably, since those energetically degenerate bound and continuum states (see the overlapped areas) belong to the subspaces of different parties, hence, hybridization cannot occur between them and yield the formation of the topological bound states in the continuum. It is observed that the corner state (bound state) with odd parity emerge in the bulk states (continuum) with even parity when $d_3$ ranges from $3.0$mm to $4.5$mm, while the corner state (bound state) with even parity emerge in the bulk states (continuum) with odd parity when $d_3$ ranges from $1.2$mm to $4.0$mm.
	
	\begin{figure*}[htbp]
		\centering\includegraphics[width=6.8in]{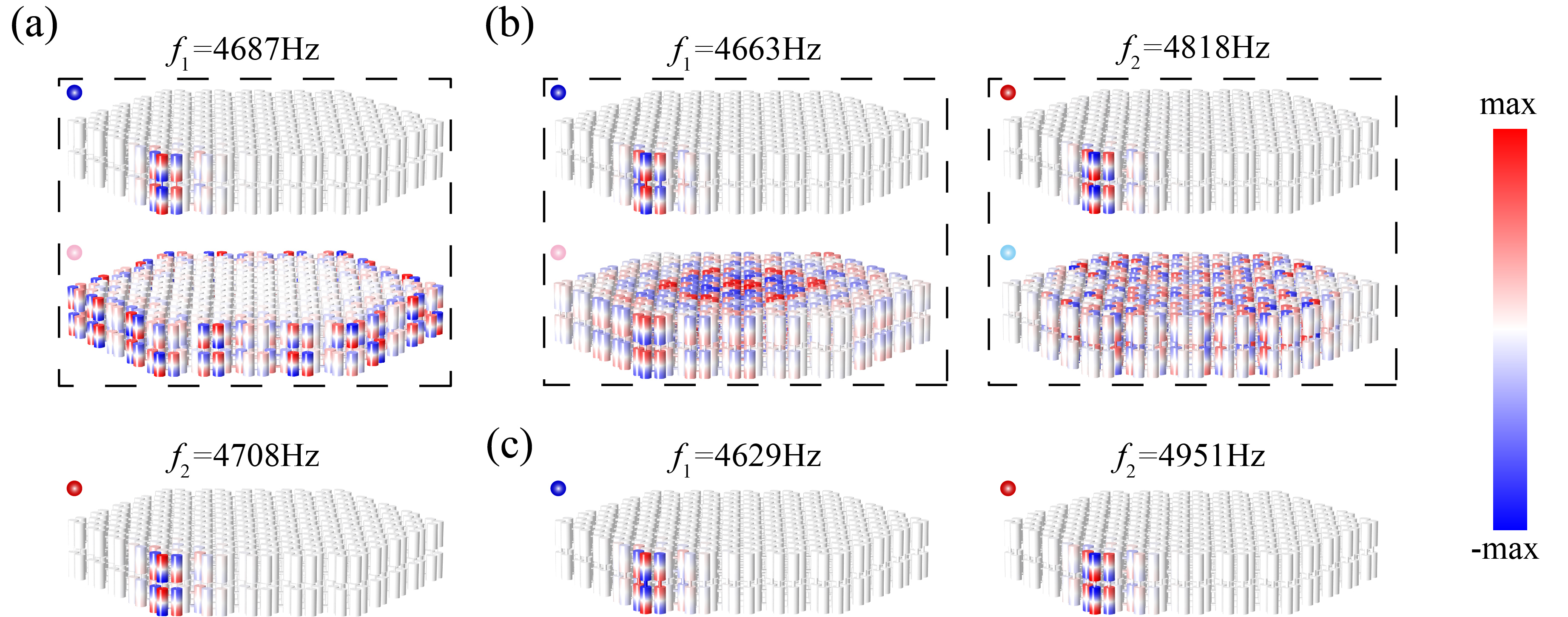}
		\caption{(a-c) The corresponding acoustic pressure field patterns of MBPC with (a) $d_3=1.2$mm at frequencies of $f_1=4687$Hz and $f_2=4708$Hz, (b) $d_3=3.0$mm at fraquencies of $f_1=4663$Hz and $f_2=4818$Hz, (c) $d_3=5.0$mm,  at frequencies of $f_1=4629$Hz and $f_2=4951$Hz, respectively. The topological bound states in the continuum can be visualized from those acoustic pressure field patterns in the dashed boxes.}
		\label{Fig_3}
	\end{figure*}
	
	To verify it, Figs.~\ref{Fig_2}(b-d) display the band structures of the MBPCs with $h_{3c}$ configuration with diameters $d_3=1.2, 3.0, 5.0$mm, respectively. As expected, accompanying the increase of the interlayer couplings, the band offset between the bandsets with even and odd mirror parities widens, and the bandset with odd (even) mirror parity move upwards (downwards). Accordingly, we further present the eigen spectra of the finite-sized MBPCs in Fig~\ref{Fig_2}(e-g). Note that the eigen states with even and odd mirror parities are separately plotted, as indicated by the blue and red points. It is seen that due to the lack of the chiral symmetry, the corner states are no longer pinned at the center of the bulk gap. As shown in Fig.~\ref{Fig_2}(e), when $d_3=1.2$mm, the frequency of corner states with even parity enter into the frequency range of the edge sates with odd parity, yielding the bound states in the edge. Meanwhile, the frequency of the corner state with odd parity is in the frequency gap of the bandset with even parity. The corresponding acoustic field patterns of two typical frequencies $f_1=4687$Hz and $f_2=4708$Hz, which topological bound states in the continuum at $f_1$ are highlighted by the dashed box, can be visualized in Fig.~\ref{Fig_3}(a). For the case of $d_3=3.0$mm in Fig.~\ref{Fig_2}(f), the frequency of the corner states with even (odd) parities move to the frequency range of bulk states with odd (even) parity, yielding the two topological bound states in the continuum, which can be visualized more clearly in Fig.~\ref{Fig_3}(b). Further increasing $d_3$ to $5.0$mm [see Fig.~\ref{Fig_2}(g)], the corner states with even (odd) parity move away from the frequency range of the bandset with odd (even) parity. Hence, only topological corner states are observed for a given freqnecy [see the acoustic field patterns of the corner states with even and odd parities in Fig.~\ref{Fig_3}(d)]. We remark that the mirror-stacking approach provides a universal way to realize topological bound states in the continuum, which have been revealed in Ref.~\cite{bilayerBIC}.
	
	\subsection{$C_6$-symmetric MBPC formed by hexagonal lattice ($h^{(6)}_{4b}$ and $h^{(6)}_{4b}$)}
	
\begin{figure}[htbp]
		\centering
		\includegraphics[width=\columnwidth]{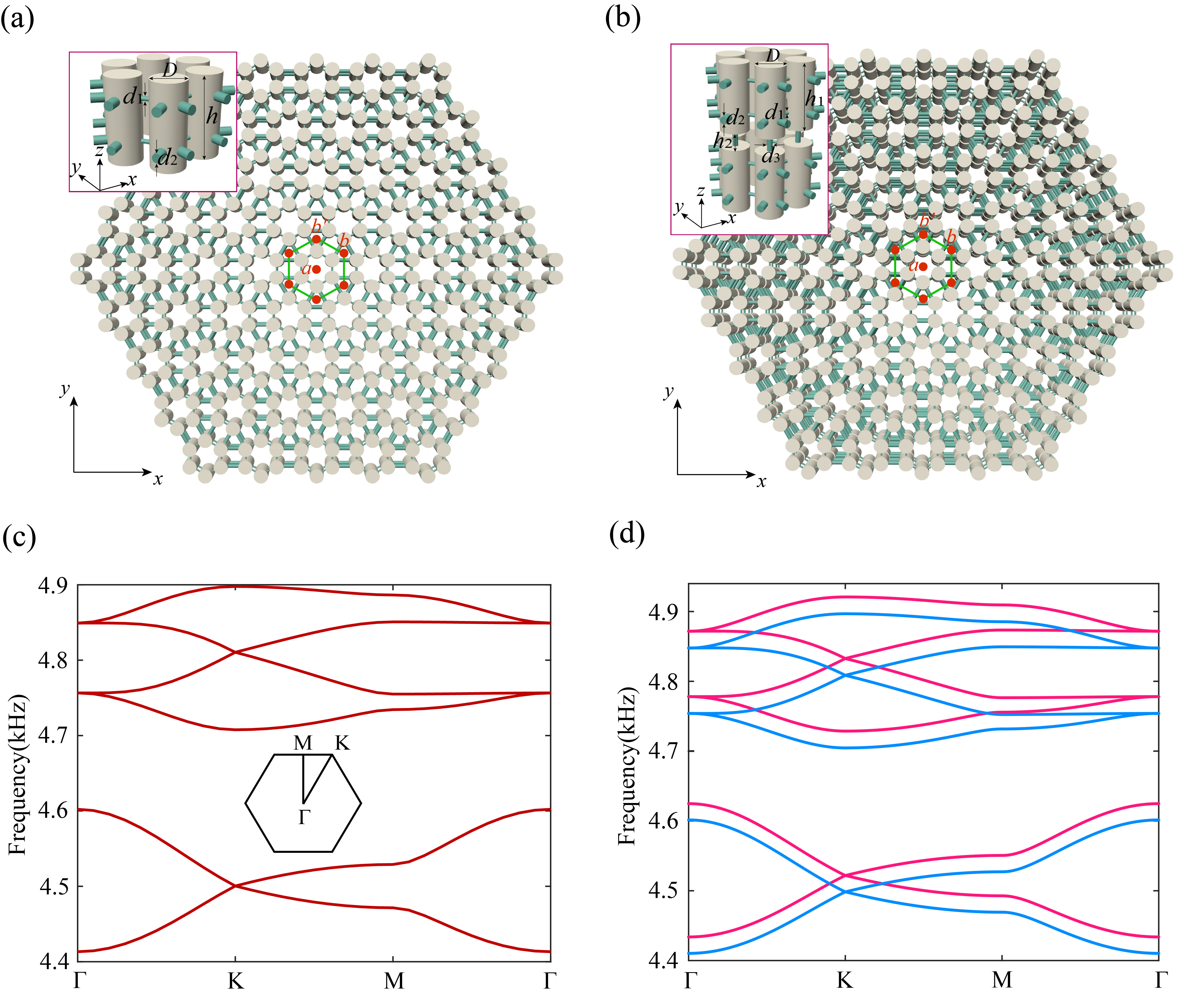}
		\caption{(a) Schematic of monolayer phononic crystals with $h^{(6)}_{4b}$ configuration. Inset: the side view of the primitive cell. (b)The MBPCs formed by two identical monolayers with $h^{(6)}_{4b}$ configuration. (c) Band structure of monolayer phononic crystals. (d) Band structure of the MBPC formed by two identical monolayers with $h^{(6)}_{4b}$ configuration.}
		\label{Fig_4}
	\end{figure}

	We then proceed to discuss the MSBC formed by stacking two identical monolayers with $h_{4b}$ configuration. As shown in Fig.~\ref{Fig_4}(a), the unit cell indicated by the green hexagon is consisting of six cavity resonators, in which each cavity resonators are placed at the middle point of the line from the corners to the center of the unit cell. For simplicity, we denote it as hexagonal lattice. Note that the topological property of the hexagonal lattice is different from that of the Wu–Hu’s lattice in spite of the similarity in the lattice configurations. To illustrate it, we keep all the geometric parameters of the monolayer phononic crystals with hexagonal lattice the same with that of the monolayer phononic crystals with Wu-Hu's lattice, namely, $H=38$mm and $D=16$mm, $d_1=2.0$mm and $d_2=4.2$mm. The band structure is depicted in Fig.~\ref{Fig_4}(c). It is seen that there are only two bands below the gap in hexagonal lattice, differ from that in Wu-Hu's lattice. Moreover, since the intercell coupling is larger than the intracell coupling ($d_2>d_1$), we remark that such a complete band gap is of a nontrivial topological nature. In the view of the Wannier configuration, the Wannier centers of the lowest two bands are located at positions $b$ and $b'$, respectively [also see the green hexagon in Fig.~\ref{Fig_4}(a)], which, in fact, is a monlayer configured with $h^{(6)}_{4b}$. 
	
	Next, we construct a MBPC by stacking the monolayer configured with $h^{(6)}_{4b}$ along $z$-direction  [Fig.~\ref{Fig_4}(b)]. The radii of the connecting tubes are remained the same with that in Fig.~\ref{Fig_1}(b).
	As shown in Fig.~\ref{Fig_4}(d), it is seen that there are twelve bands, which according to their mirror parities, can be divided into two sets. The blue (red) bandset refers to the bands with even (odd) mirror parity. Obviously, the interlayer coupling results in the splitting of the bands with odd and even parities.
	
	\begin{figure*}[htbp]
		\centering
		\includegraphics[width=6.8in]{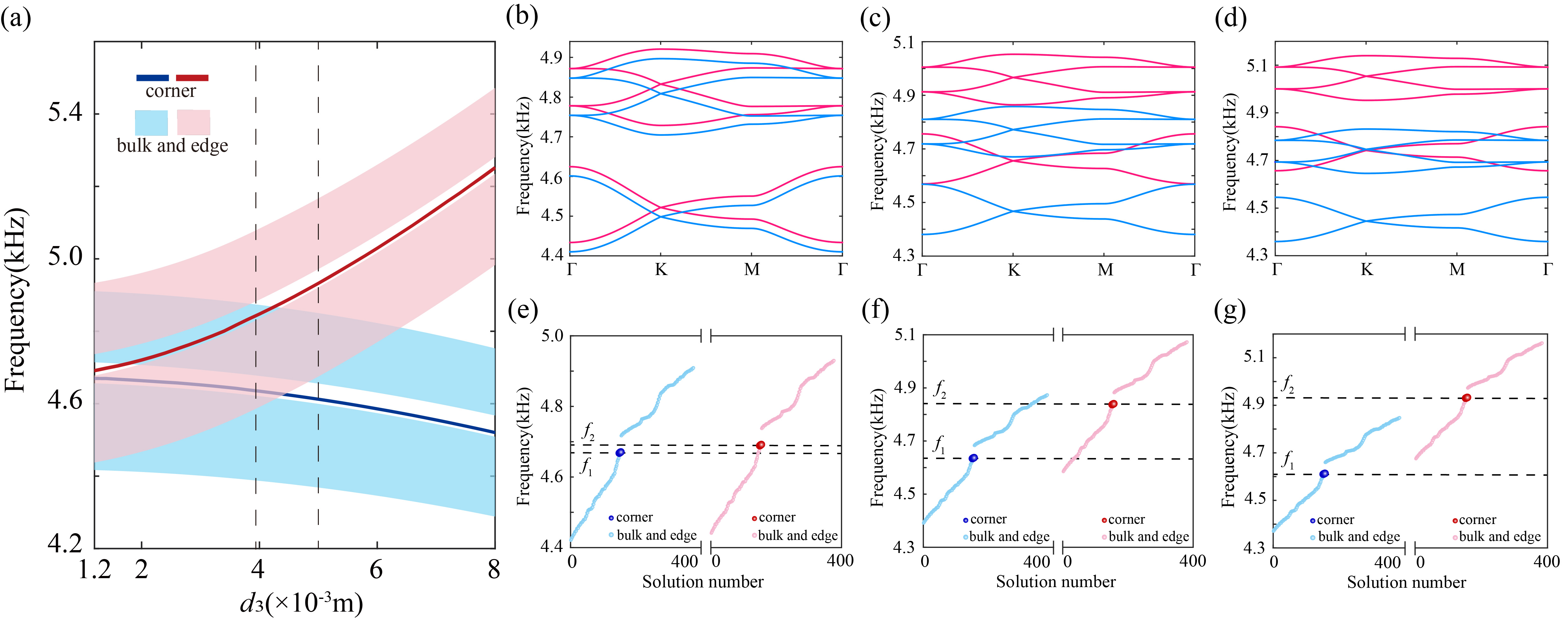}
		\caption{(a) The eigen spectra of MBPC versus the interlayer coupling $d_3$. (b-d) The band structure of the mirror-stacked bilayer phononic crystals with interlayer coupling (b) $d_3=1.2$mm, (c) $d_3=3.9$mm, (d) $d_3=5.0$mm. (e-g) The corresponding of eigen spectra of the MBPC in (b-d), where the blue and red points refer to the acoustic states with even and odd mirror parities, respectively.}
		\label{Fig_5}
	\end{figure*}
	
	\begin{figure*}[htbp]
		\centering
		\includegraphics[width=6.8in]{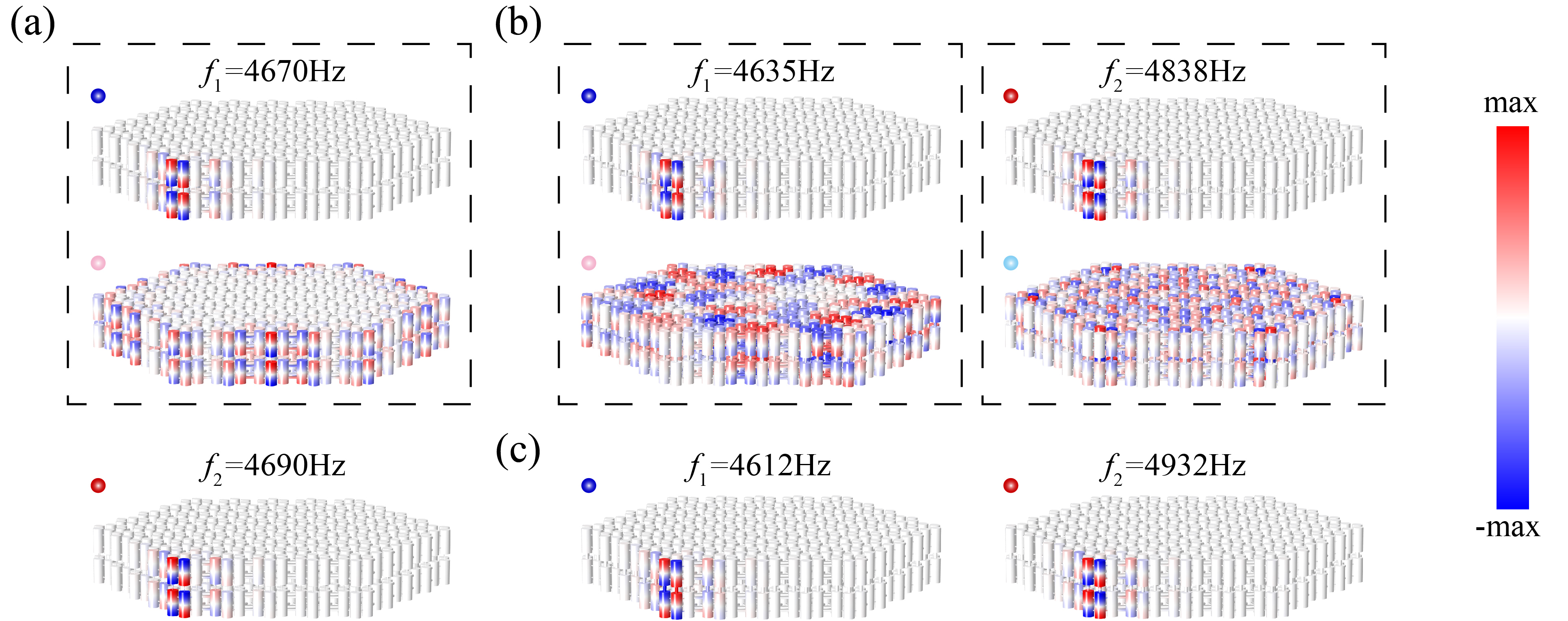}
		\caption{(a-c) The corresponding acoustic pressure field pattern of MBPC with (a) $d_3=1.2$mm at fraquencies of $f_1=4670$Hz and $f_2=4690$Hz, (b) $d_3=3.9$mm at fraquencies of $f_1=4635$Hz and $f_2=4838$Hz, (c) $d_3=5.0$mm,  at fraquencies of $f_1=4612$Hz and $f_2=4932$Hz, respectively. The topological bound states in the continuum can be visualized from those acoustic pressure field patterns in the dashed boxes.}
		\label{Fig_6}
	\end{figure*}
	
	Owing to the higher-order band topology in the monolayer with $h^{(6)}_{4b}$ configuration, it is evident that each layer structure with finite size supports six corner states in the gap. Hence, when two monolayers with $h^{(6)}_{4b}$ configuration are placed together via mirror-stacking approach, it is expected that each layer of MBPCs hosts corner states inheriting from higher-order band topology of the monolayer. Similar to the bulk band with specific mirror parity, twelve corner states in the gap are reorganized into two groups according to the mirror parities. Therefore, it is possible that the corner states can move continuously into and out of the two-dimensional bulk continuum of opposite parity by tuning the interlayer couplings, which results in the appearance and disappearance of the topological bound states in the continuum. 
	
	To this end, we plot the energy spectra of the finite-sized MBPCs versus the diameter $d_3$ of the connecting air tube in Fig.~\ref{Fig_5}(a). The shadow light blue (red) area and blue (red) line refer to the bulk and corner states of MBPC with even (odd) parity, respectively. For each subspace, it is clearly seen that a corner state emerges as the manifestation of the higher-order band topology. Specifically, for the diameter $d_3$ ranges from $2.0$mm ($1.2$mm) to $4.2$mm ($4.4$mm), the frequency of the corner states with odd (even) parity falls into the frequency windows of the bulk states with even (odd) parity, giving rise to the topological bound states in the continuum. Since those energetically degenerate bound and continuum states belong to the subspaces of different parties, hence, hybridization cannot occur between them and yield the formation of the topological bound states in the continuum. 
	
	Moreover, Figs.~\ref{Fig_5}(b-d) display the band structures of the MBPCs formed by two hexagonal lattice with diameters $d_3=1.2, 3.9, 5.0$mm, respectively. As expected, the band offset between the bandsets with even and odd mirror parities widens, which the bandset with odd (even) mirror parity move upwards (downwards) accompanying the increase of the interlayer couplings. Accordingly, we present the eigen spectra of finite-sized  MBPCs in Fig.~\ref{Fig_5}(e-g). Note that the eigen states with even and odd mirror parities are separately plotted, as indicated by the blue and red points. It is seen that due to the lack of the chiral symmetry, the corner states no longer pinned at the center of the bulk gap. As shown in Fig.~\ref{Fig_5}(e), when $d_3=1.2$mm, the frequency of corner states with even parity enter into the frequency range of the edge sates with odd parity, yielding the bound states in the edge [also see the corresponding acoustic field pattern at a frequency of $f_1=4670$Hz in the dashed box of Fig.~\ref{Fig_6}(a)]. Meanwhile, the frequency of the corner state with odd parity is in the frequency gap of the bandset with even parity [also see the corresponding acoustic field pattern at a frequency of $f_1=4690$Hz in Fig.~\ref{Fig_6}(a)]. For the case of $d_3=3.9$mm in Fig.~\ref{Fig_5}(f), the frequency of the corner states with even (odd) parities, namely, $f_1=4635$Hz ($f_2=4838$Hz), fall into the frequency range of bulk states with odd (even) parity, yielding the topological bound states in the continuum, which can be visualized more clearly from the acoustic field patterns in the dashed box in Fig.~\ref{Fig_6}(b). Further increasing $d_3$ to $5.0$mm [see Fig.~\ref{Fig_5}(g)], the frequency of the corner states with even (odd) parity, saying, $f_1=4612$Hz ($f_2=4932$Hz), is in the frequency gap of the bandset with odd (even) parity [see the acoustic field pattern of the corner states with even and odd parities in Fig.~\ref{Fig_6}(d)]. It is believed that the mirror-stacking approach provides a universal way to realize topological bound states in the continuum.
	
	\subsection{$C_3$-symmetric MBPC formed by kagome lattice ($h^{(3)}_{2b}$ and $h^{(3)}_{2b}$)}
	
\begin{figure}[htbp]
		\centering
		\includegraphics[width=\columnwidth]{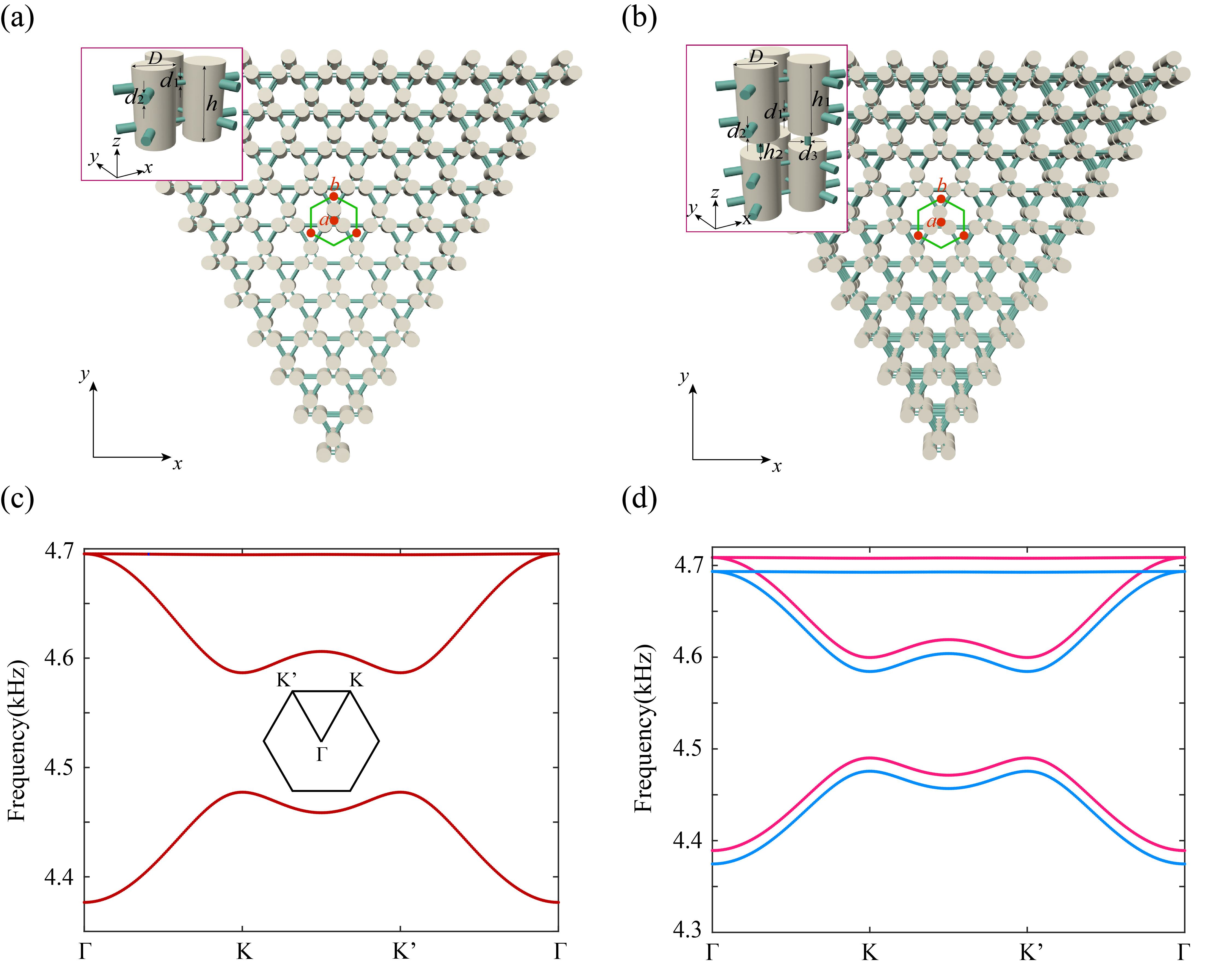}
		\caption{(a) Schematic of monolayer phononic crystals with $h^{(3)}_{2b}$ configuration. Inset: the side view of the unit cell. (b) The MBPC formed by two identical monolayers with $h^{(3)}_{2b}$ configuration. (c) Band structure of monolayer phononic crystals. (d) Band structure of the MBPC with $h^{(3)}_{2b}$ configuration.}
		\label{Fig_7}
	\end{figure}

	Following the above procedures, we move to discuss MBPCs formed by stacking monolayer phononic crystals with $C_3$ symmetry. Typical crystalline insulators with $C_3$ symmetry is the well-known kagome lattices. Hence, we start with the MSBC formed by stacking two identical monolayers made up by kagome lattice. As shown in Fig.~\ref{Fig_7}(a), each unit cell made up by three cavity resonators, which are aligned with upward triangular shape. For simplicity, we denoted as upward kagome lattice. The geometric parameters of the unit cell are as follows: the diameters of the connecting air tube within the unit cell is $d_1=2.4$mm, and that between two nearby unit cells is $d_2=4.6$mm. The height and diameter of the individual cavity is $h_1=38$mm and $D=20$mm. Note that such a geometry configuration is equivalent to a tight-binding model that intercell coupling is larger than the intracell coupling. We further construct a MBPC by stacking the monolayer with upward kagome lattice along $z$-direction, as indicated in Fig.~\ref{Fig_7}(b). The diameter of the connecting tube between two layers is given by $d_3$, which is originally set as $1.2$mm. 
	
	Furthermore, we present the band structure of the monolayer with upaward kagome lattice in Fig.~\ref{Fig_7}(c). It is seen that only a single band below the band gap. Since the intercell coupling is larger than the intracell coupling ($d_2>d_1$),  we remark that such a complete band gap is of a nontrivial topological nature according to Ref.~\cite{phys.rev.lett.120.026801ezawa}. In view of the Wannier configuration, the Wannier center of the lowest band are located at position $b$ [also see the green hexagon in Fig.~\ref{Fig_7}(a)]. Hence, such a monolayer is of $h^{(3)}_{2b}$ configuration. Figure~\ref{Fig_7}(d) depicts the band structure of the MSBC made up by two monolayers with $h^{(3)}_{2b}$ configuration. It is seen that there are six bands, which according to their mirror parities, can be divided into two sets. The blue (red) bandset refers to the bands with even (odd) mirror parity. Obviously, the interlayer coupling results in the splitting of the bands with odd and even parities.
	
	\begin{figure*}[htbp]
		\centering\includegraphics[width=6.8in]{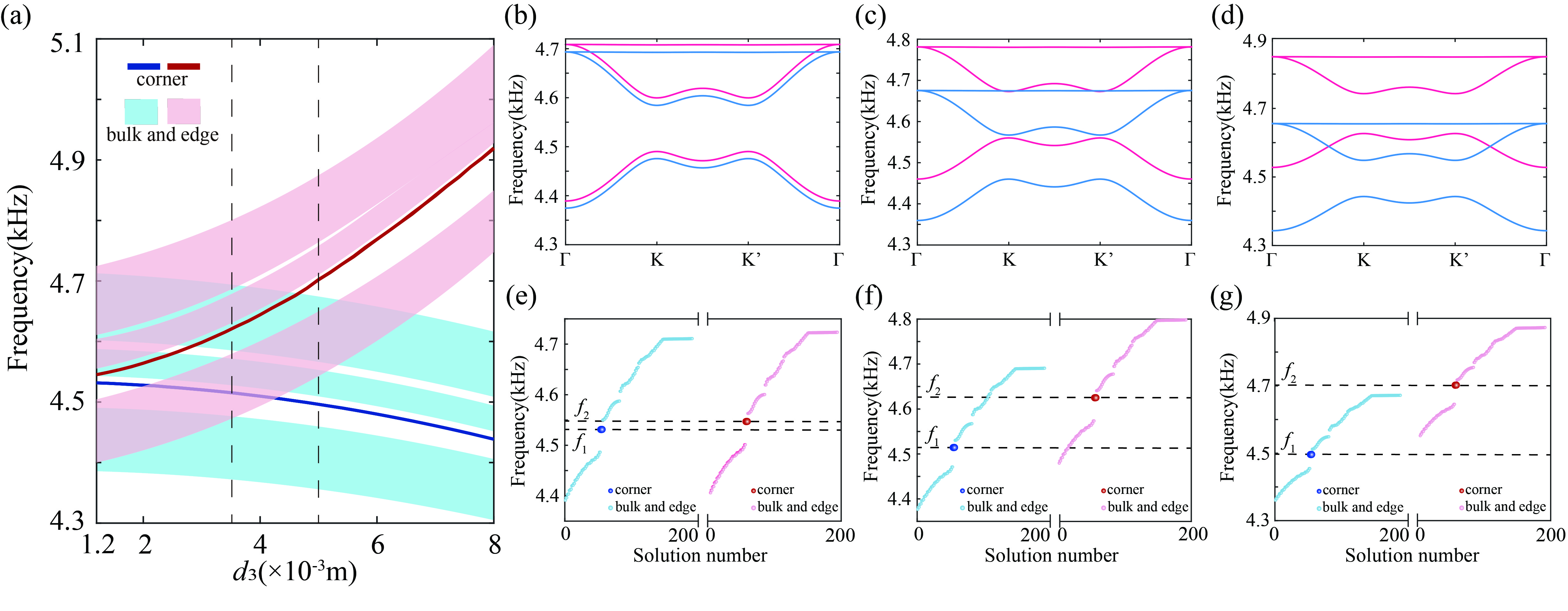}
		\caption{(a) The eigen spectra of MBPC versus the diameter $d_3$ of the layer connecting tubes. (b-d) The band structure of the MBPCs with the diameter $d_3$ of the layer connecting tubes (b) $d_3=1.2$mm, (c) $d_3=3.6$mm, (d) $d_3=5.0$mm. (e-g) The corresponding of eigen spectra of the MBPC in (b-d), where the blue and red points refer to the acoustic states with even and odd mirror parities, respectively.}
		\label{Fig_8}
	\end{figure*}
	
	\begin{figure*}[htbp]
		\centering\includegraphics[width=6.8in]{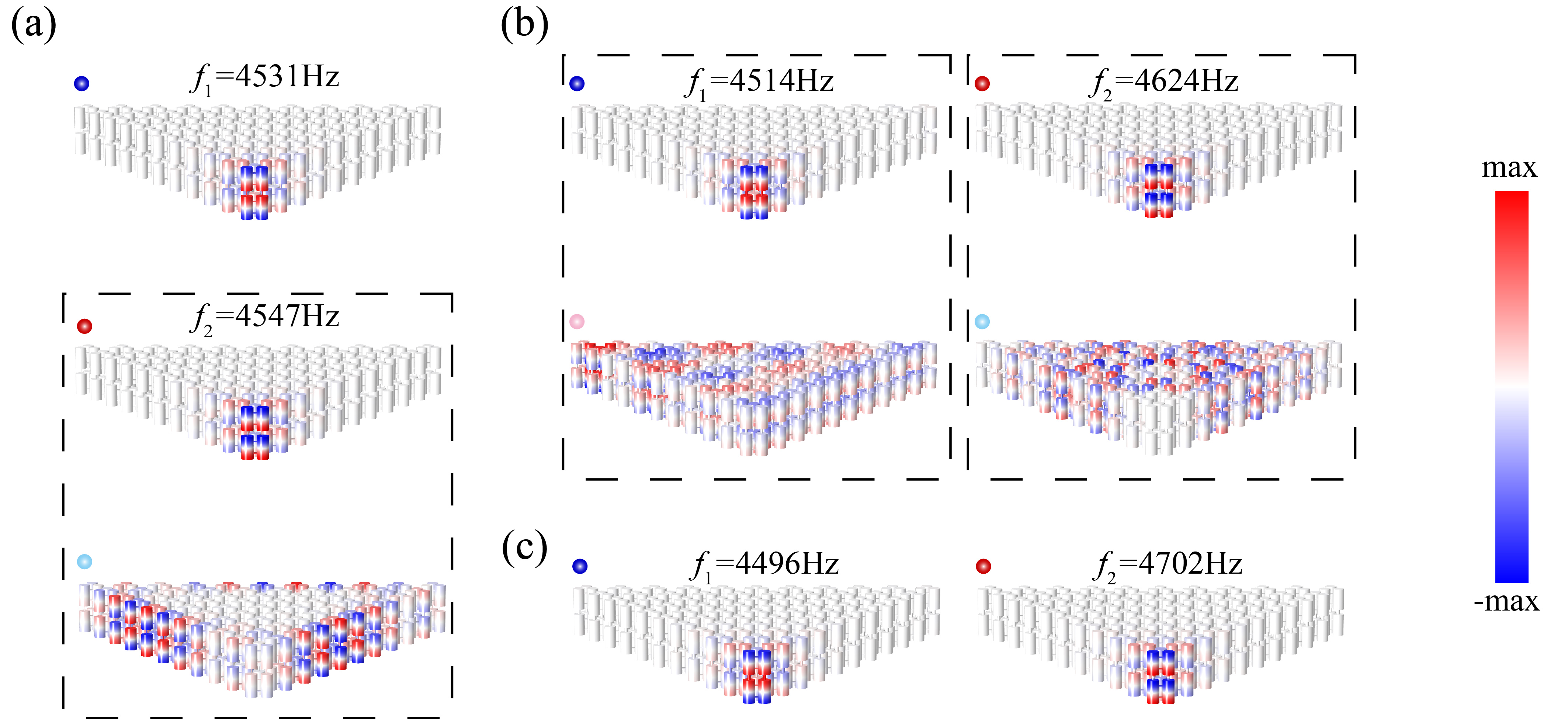}
		\caption{(a-c) The corresponding acoustic pressure field pattern of MBPC with (a) $d_3=1.2$mm at fraquencies of $f_1=4531$Hz and $f_2=4547$Hz, (b) $d_3=3.6$mm at fraquencies of $f_1=4514$Hz and $f_2=4624$Hz, (c) $d_3=5.0$mm,  at fraquencies of $f_1=4496$Hz and $f_2=4702$Hz, respectively. The topological bound states in the continuum can be visalized from those acoustic pressure field patterns in the boxes.}
		\label{Fig_9}
	\end{figure*}
	
	Owing to the higher-order band topology in the monolayer, it is evident that each layer structure with finite size supports three corner states in the gap. When two monolayer are placed together via mirror-stacking approach, each layer also hosts corner states that inherit from higher-order band topology of the monolayer. Similar to the bulk band with specific mirror parity, six corner states in the gap are reorganized into two groups according to their mirror parities. Therefore, it is possible that the corner states can move continuously into and out of the two-dimensional bulk continuum of opposite parity by tuning the interlayer couplings, which contribute a topological bound states in the continuum.
	
	To this end, we plot the energy spectra of the finite-sized MBPC made up by monolayer with $h^{(3)}_{2b}$ versus the diameter of the interlayer connecting air tubes in Fig.~\ref{Fig_8}(a). The shadow light blue (red) area and blue (red) line refers to the bulk and corner states of MBPC with even (odd) parity, respectively. For each subspace, it is clearly seen that the corner states emerge as the manifestation of the higher-order band topology. Moreover, the corner states with odd (even) parity fall into the frequency windows of the bulk states with even (odd) parity when the diameter $d_3$ range from 1.2mm (2.0mm) to 4.6mm (4.2mm). Remarkably, since those energetically degenerate bound and continuum states (see the overlapped areas) belong to the subspaces of different parties, hence, hybridization cannot occur between them and yield the formation of the topological bound states in the continuum. 
	
	To verify it, Figs.~\ref{Fig_8}(b-d) display the band structures of the MBPCs with various diameters $d_3=1.2, 3.6, 5.0$mm, respectively. As expected, the band offset between the bandsets with even and odd mirror parities widens, which the bandset with odd (even) mirror parity move upwards (downwards) accompanying the increase of the interlayer couplings. Accordingly, we further present the eigen spectra of MBPC with finite-sized systems in Fig~\ref{Fig_8}(e-g). Note that the eigen states with even and odd mirror parities are separately plotted, as indicated by the blue and red points. It is seen that due to the lack of the chiral symmetry, the corner states are no longer pinned at the center of the bulk gap. As shown in Fig.~\ref{Fig_8}(e), when $d_3=1.2$mm, the corner states at a frequency of $f_2=4547$Hz with odd parity enter into the frequency range of the edge sates with even parity, yielding the bound states in the edge. Meanwhile, the frequency of the corner state at a frequency of $f_1=4531$Hz with even parity is in the frequency gap of the bandset with odd parity. The corresponding acoustic field pattern can be visualized in Fig.~\ref{Fig_9}(a). For $d_3=3.6$mm in Fig.~\ref{Fig_8}(f), the corner states at a frequency of $f_1=4514$Hz ($f_2=4624$Hz) with even (odd) parities move to the frequency range of bulk states with odd (even) parity, yielding the topological bound states in the continuum, which can be visualized more clearly in the dashed boxes in Fig.~\ref{Fig_9}(b). Further increasing $d_3$ to $5.0$mm [see the eigen spectrum Fig.~\ref{Fig_8}(g)], the corner states at a frequency of $f_1=4496$Hz ($f_2=4702$Hz) with even (odd) parity move away from the frequency windows of the bandset with odd (even) parity, leading to the disappear of the topological bound states in the continuum [see the acoustic field pattern of the corner states with even and odd parities in Fig.~\ref{Fig_9}(c)]. 
	
	\subsection{$C_3$-symmetric MBPC formed by kagome lattice ($h^{(3)}_{2c}$ and $h^{(3)}_{2c}$)}
	
\begin{figure}[htbp]
		\centering
		\includegraphics[width=\columnwidth]{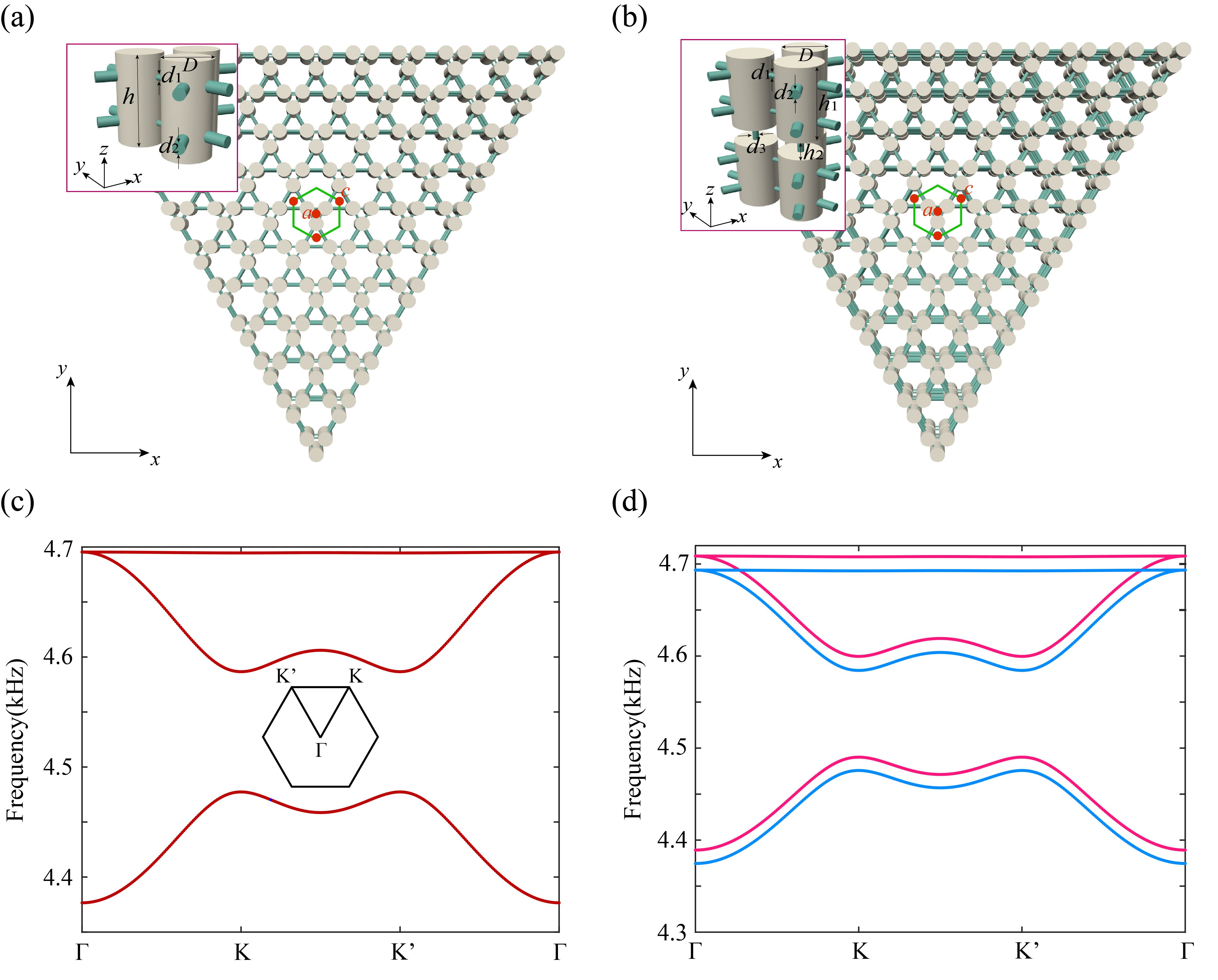}
		\caption{(a) Schematic of monolayer made up by the downward kagome lattice. Inset: the side view of the primitive cell. (b) The MBPC formed by two identical monolayers with $h^{(3)}_{2c}$ configuration. (c) Band structure of monolayer phononic crystals. (d) Band structure of the MBPC with $h^{(3)}_{2c}$ configuration.}
		\label{Fig_10}
	\end{figure}

	It is note worthing that there exists another $C_3$-symmetric lattice. As shown in Fig.~\ref{Fig_10}(a), each unit cell made up by three cavity resonators, which are aligned with downward triangular shape. For simplicity, we denoted as downward kagome lattice. Note that the geometric parameters of the downward kagome lattice are same with that of upward kagome lattice. Hence, the downard kagome lattice and upward kagome lattice are mirror symmetric with respect to each. Moreover, it is evident that the monolayer phononic crystals arrayed in downward kaogme lattice share the same band structure with that in upward kaogme lattice, which is depicted in Fig.~\ref{Fig_10}(c). From the band diagram, it is seen that only a single band below the band gap. Since the intercell coupling is larger than the intracell coupling ($d_2>d_1$), we remark that such a complete band gap is of a nontrivial topological nature according to Ref.~\cite{phys.rev.lett.120.026801ezawa}. Nevertheless, we also remark that the topological index of downward kagome lattice is different from that of upward kagome lattice (with $h_{2b}$ configuration). In view of the Wannier configuration, the Wannier center of the lowest band are located at position $c$ [also see the green hexagon in Fig.~\ref{Fig_10}(a)]. Hence, the monolayer with downward kagome lattice is of $h^{(3)}_{2c}$ configuration.
	
	Next, we further construct a $C_3$-symmetric MBPC with two stacked downward kagome lattice, as indicated in Fig.~\ref{Fig_10}(b). The diameter of the connecting air tube between two layers is given by $d_3$. Figure~\ref{Fig_10}(d) depicts the band structure of the MSBC. As expected, they share the identical band structure of the monolayer and MSBCs with $h_{2b}$ configuration. It is seen that there are six bands, which according to their mirror parities, can be divided into two sets. The blue (red) bandset refers to the bands with even (odd) mirror parity. Obviously, the interlayer coupling results in the splitting of the bands with odd and even parities.
	
	\begin{figure*}[htbp]
		\centering
		\includegraphics[width=6.8in]{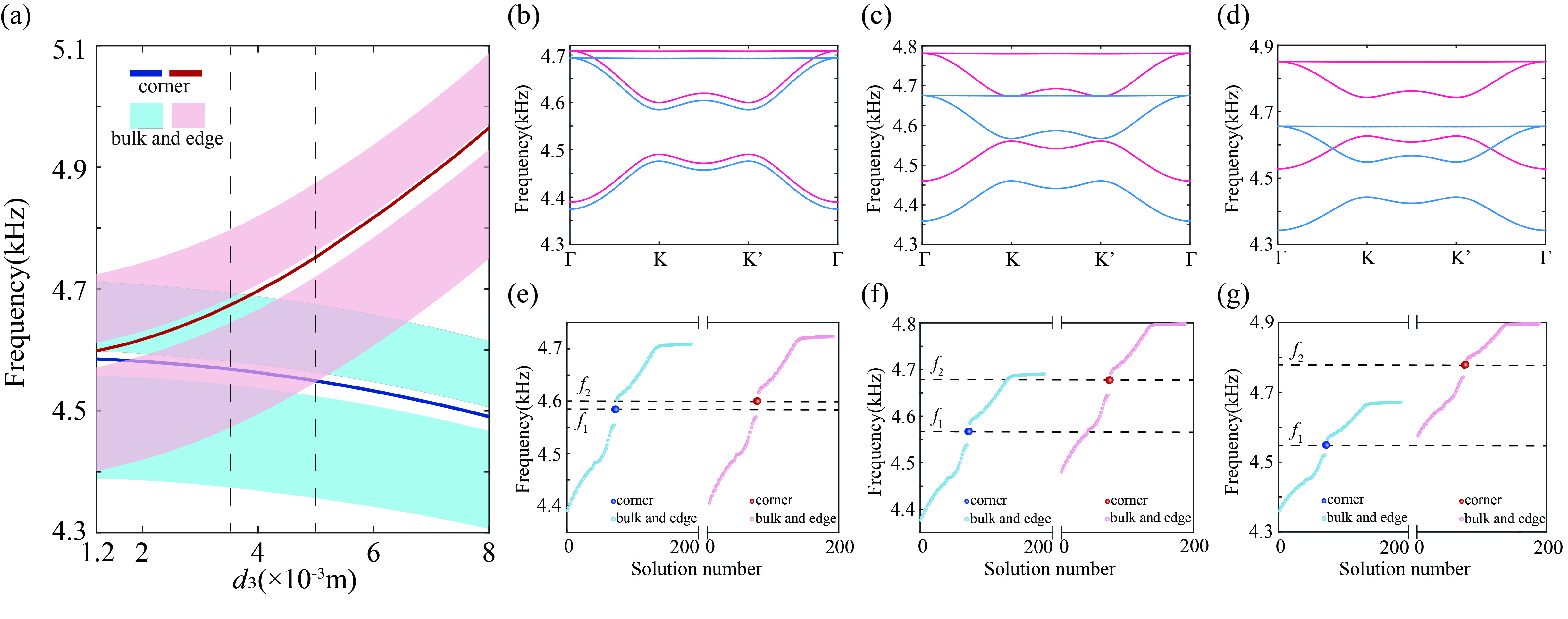}
		\caption{(a) The eigen spectra of MBPC versus the interlayer coupling $d_3$. (b-d) The band structure of the MBPCs with interlayer coupling (b) $d_3=1.2$mm, (c) $d_3=3.6$mm, (d) $d_3=5.0$mm. (e-g) The corresponding of eigen spectra of the mirror-stacked bilayer phononic crystal in (b-d), where the blue and red points refer to the acoustic states with even and odd mirror parities, respectively.}
		\label{Fig_11}
	\end{figure*}
	
	\begin{figure*}[htbp]
		\centering
		\includegraphics[width=6.8in]{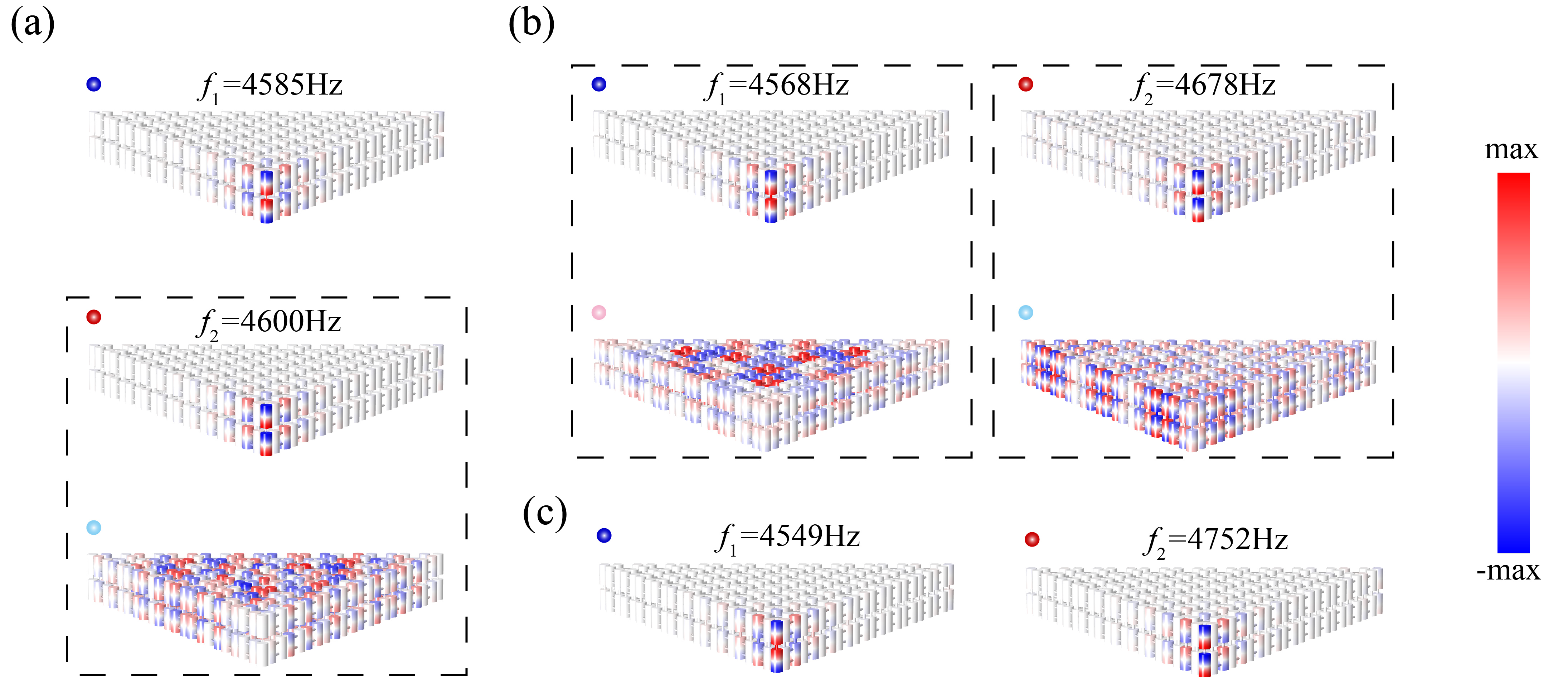}
		\caption{(a-c) The corresponding acoustic pressure field pattern of MBPC with (a) $d_3=1.2$mm at fraquencies of $f_1=4585$Hz and $f_2=4600$Hz, (b) $d_3=3.6$mm at fraquencies of $f_1=4568$Hz and $f_2=4678$Hz, (c) $d_3=5.0$mm,  at fraquencies of $f_1=4549$Hz and $f_2=4752$Hz, respectively. The topological bound states in the continuum can be visalized from those acoustic pressure field patterns in the boxes.}
		\label{Fig_12}
	\end{figure*}
	
	Owing to the higher-order band topology in the monolayer, it is evident that each layer structure with finite size supports three corner states in the gap. When two monolayers are placed together via mirror-stacking approach, each layer also hosts corner states that inherit from higher-order band topology of the monolayer. Similar to the bulk band with specific mirror parity, six corner states in the gap are reorganized into two groups according to the mirror parities. Therefore, it is possible that the corner states can move continuously into and out of the two-dimensional bulk continuum of opposite parity by tuning the interlayer couplings, which contribute a topological bound states in the continuum. 
	
	To this end, we plot the energy spectra of finite-sized MBPC versus the diameter $d_3$ of the interlayer connecting air tube in Fig.~\ref{Fig_11}(a). The shadow light blue (red) area and the blue (red) line refer to the bulk and corner states of MBPC with even (odd) parity, respectively. For each subspace, it is clearly seen that a corner state emerges as the manifestation of the higher-order band topology. Moreover, the corner states with odd (even) parity fall into the frequency windows of the bulk states with even (odd) parity when the diameter $d_3$ ranges from 1.2mm (1.6mm) to 3.8mm (5.0mm). Remarkably, since those energetically degenerate bound and continuum states (see the overlapped areas) belong to the subspaces of different parties, hence, hybridization cannot occur between them and yielding the formation of the topological bound states in the continuum. 
	
	To verify it, Figs~\ref{Fig_11}(b-d) display the band structures of the MBPCs with various diameters $d_3=1.2, 3.6, 5.0$mm, respectively. As expected, the band offset between the bandset with even and odd mirror parities widened, which the bandset with odd (even) mirror parity move upwards (downwards) accompanying the increase of the interlayer couplings. Accordingly, we further present the eigen spectra of finite-sized MBPC in Fig~\ref{Fig_12}(e-g). Note that the eigen states with even and odd mirror parities are separately plotted, as indicated by the blue and red points. It is seen that due to the lack of the chiral symmetry, the corner states no longer pinned at the center of the bulk gap. As shown in Fig.~\ref{Fig_11}(e), when $d_3=1.2$mm, the corner states at a frequency of $f_2=4600$Hz with odd parity fall into the frequency range of the bulk sates with even parity, yielding the bound states in the continuum. Meanwhile, the corner states at a frequency of $f_1=4585$Hz with even parity is in the frequency gap of the bandset with odd parity. The corresponding acoustic field pattern can be visualized in Fig.~\ref{Fig_12}(a). For the case of $d_3=3.6$mm in Fig.~\ref{Fig_11}(f), the corner states at a frequency of $f_2=4678$Hz ($f_1=4568$) with odd (even) parity fall into the frequency range of bulk states with even (odd) parity, yielding the topological bound states in the continuum, which can be visualized more clearly in the dashed boxes of Fig.~\ref{Fig_12}(b). Further increasing $d_3$ to $5.0$mm [see Fig.~\ref{Fig_11}(g)], the corner states at a frequency of $f_2=4752$Hz ($f_1=4549$Hz) with odd (even) parity move into the frequency gap of the banset with even (odd) parity [see the acoustic field patterns in Fig.~\ref{Fig_12}(c)].
	
	To conclude this section, we emphasize that the interlayer coupling controlled by the diameter of the connecting air tubes in the MBPCs plays a vital role on the emergence of topological bound states in the continuum, which the topological corner and bulk states served as bound and continuum states, respectively. To be specific, the mirror symmetry enables the separation of the Hilbert space and hence hinders the hybridization of the the topological corner and bulk states with different mirror symmetry eigenvalues. The weak interlayer coupling give rise to the emergence of the topological bound states in the continuum, while the large interlayer coupling makes the topological bound states in the continuum disappear.
	
	\section{Heterogeneous-stacked bilayer phononic crystals}
	
	Based on the above studies, we further discuss the HSBCs formed by stacking two different $C_6$ or $C_3$-symmetric monolayers phononic crystals in this section.  
	
	\subsection{$C_6$-symmetric HBPC formed by Wu-Hu's and hexagonal lattices ($h^{(6)}_{3c}$ and $h^{(6)}_{4b}$)} 
	
\begin{figure*}[htbp]
		\centering\includegraphics[width=6.8in]{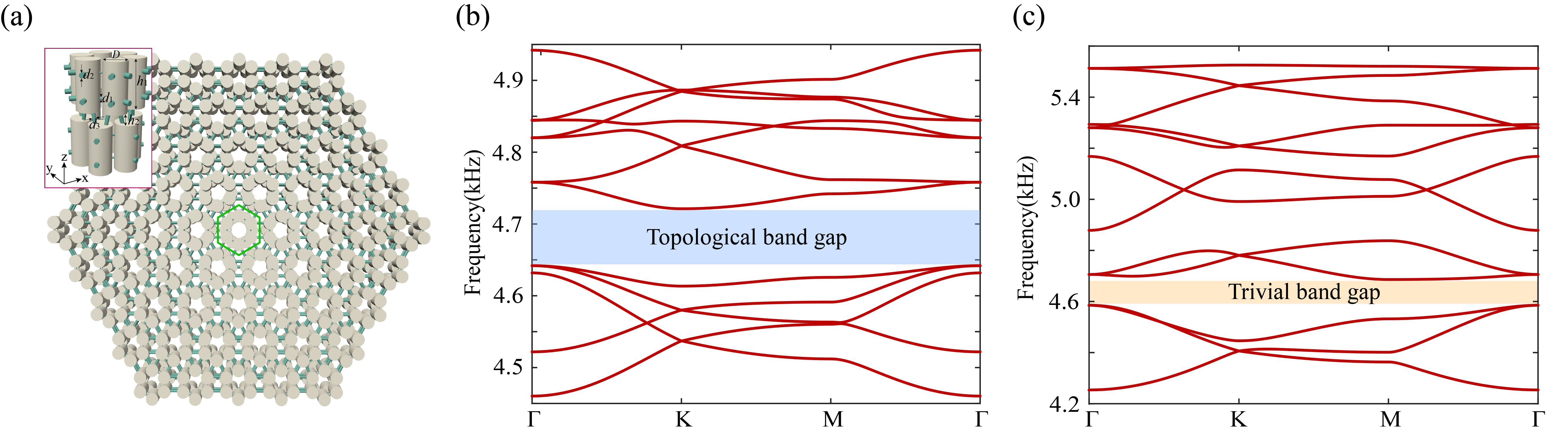}
		\caption{ (a) Top view of the $C_6$-symmetric HBPC, which formed by connecting the monolayers phononic crystal arrayed with Wu-Hu's and hexagon lattices ($h^{(6)}_{4b}$ and $h^{(6)}_{3c}$), respectively. Inset: the unit cell. The geometric parameters of two monolayer phononic crystals are adopted from Figs.~\ref{Fig_1}(a) and ~\ref{Fig_4}(a), respectively. The diameter of the individual cavity is $D=18$mm (b,c) Simulated band structure of HBPC with (b) weak interlayer coupling, i.e., the diameter of the connecting air tube $d_3=1.2$ mm, and (c) strong interlayer coupling, i.e., the diameter of the connecting air tube $d_3=8.0$ mm.}
		\label{Fig_13}
	\end{figure*}

	We first consider the $C_6$-symmetric HBPC formed by stacking Wu-Hu's lattice and hexagonal lattice with nontrivial band topology, i.e., configured with $h^{(6)}_{3c}$ and $h^{(6)}_{4b}$ types. To this end, we adopt the monolayer phononic crystals configured with $h^{(6)}_{3c}$ and $h^{(6)}_{4b}$, which the geometric parameters are same with the Figs.~\ref{Fig_1}(a) and ~\ref{Fig_4}(a), respectively. As shown in Fig.~\ref{Fig_13}(a), the unit cell indicated by the green hexagon consisting of twelve acoustic cavities, each cavity is connected with two nearest neighbour cavities (within the unit cell) of upper or lower layer via two sloped air tubes. We then study the effect of interlayer coupling on the band topology in HBPCs with different interlayer couplings. As shown in Figs.~\ref{Fig_13}(b) and ~\ref{Fig_13}(c), we present the band structure of the HBPC with $d_3=1.2$mm and $d_3=8$mm, which represent the weak and strong interlayer couplings, respectively. Under the weak interlayer coupling, namely $d_3=1.2$mm, it is seen that a band gap divide the twelve bands into two sets, the lower bandset below the gap consisting of five bands, while higher bandset above the gap consisting of seven bands. Note that the lower bandset are inherited from the band structure of the monolayers configured with $h^{(6)}_{4b}$ and $h^{(6)}_{3c}$ types, which are almost not affected. Nevertheless, due to the absence of the mirror symmetry, the separability into subspaces with opposite parities is no longer valid in HBPCs, which is in strong contrast to that in the $C_6$-symmetric MBPC. The band mixing effect originated from interlayer coupling can be further examined from the band anticrossing in the higher bandset. 
	
	On the other hand, the band gap of the monolayer phononic crystal with $h^{(6)}_{4b}$ and $h^{(6)}_{3c}$ configuration share a common frequency range, which, however, are of different nontrivial topology. Hence, we further employ the topological crystalline index to characterize the band topology in HBPCs. Following Ref.~\cite{topocharater1}, the topological crystalline index can be expressed by the full set of the $C_6$ eigenvalues at the high-symmetry points (HSPs). For an HSP denoted by the symbol $\Pi$, the $C_6$ eigenvalues can only be $\Pi_n=e^{i2\pi(n-1)/6}$ with $n$ ranges from 1 to 6. Here, the HSPs include the $\Gamma$, $M$ and $K$. The minimum set of indices that describe the band topology of $C_6$ crystalline insulator is given by 
	\begin{equation}
		\begin{aligned}
			& \left[ M^{(2)}_1 \right] = \# M^{(2)}_1 -\# \Gamma^{(2)}_1,\\
			& \left[ K^{(3)}_1 \right] = \# K^{(2)}_1 -\# \Gamma^{(3)}_1,
		\end{aligned}
	\end{equation}
	where $\# M^{(2)}_1 (\# \Gamma^{(2)}_1)$ refers to the number of bands below the band gap with the $C_2$ symmetry eigenvalue $M_1 (\Gamma_1)$ at the $M (\Gamma)$ point, and $\# K^{(3)}_1 (\# \Gamma^{(3)}_1)$ refers to the number of bands below the band gap with the $C_3$ symmetry eigenvalue $K_1 (\Gamma_1)$ at the $K (\Gamma)$ point. Hence, the property of the gap can be characterized by 
	\begin{equation}
		\xi^{(6)}=\left ( \left [ M_1^{(2)}\right ], \left [K^{(3)}_1 \right ] \right ).
	\end{equation}
	The calculation show the topological index of the band gap is $(2,2)$, indicating a nontrivial higher-order topology. Hence, it is emphasized that the band topology of the HBPC inherits from the nontrivial monolayer phononic crystal under weak interlayer coupling. 
	
	We proceed to discuss the band structure of the HBPC under the strong interlayer coupling, namely $d_3=8$mm, in Fig.~\ref{Fig_13}(c). It is seen that two band gap divide the twelve bands into three sets. To be specific, the lowest bandset (with three bands) inherits the band structure of the monolayer with $h^{(6)}_{3c}$ configuration, while higher bandsets are reorganized owing to the strong interlayer coupling, which makes a difference from that weak interlayer coupling. Here we only focus on the first bandgap indicated by the yellow zone in Fig.~\ref{Fig_13}(c). Remarkably, the calculated topological crystalline index gives $\xi^{(6)}=(0,0)$, indicating a trivial phase. Hence, it is emphasized that the phase transition in the HBPC can be triggered by adjusting the interlayer couplings, rather than breaking the geometric symmetry.
	
\begin{figure*}[htbp]
		\centering\includegraphics[width=6.8in]{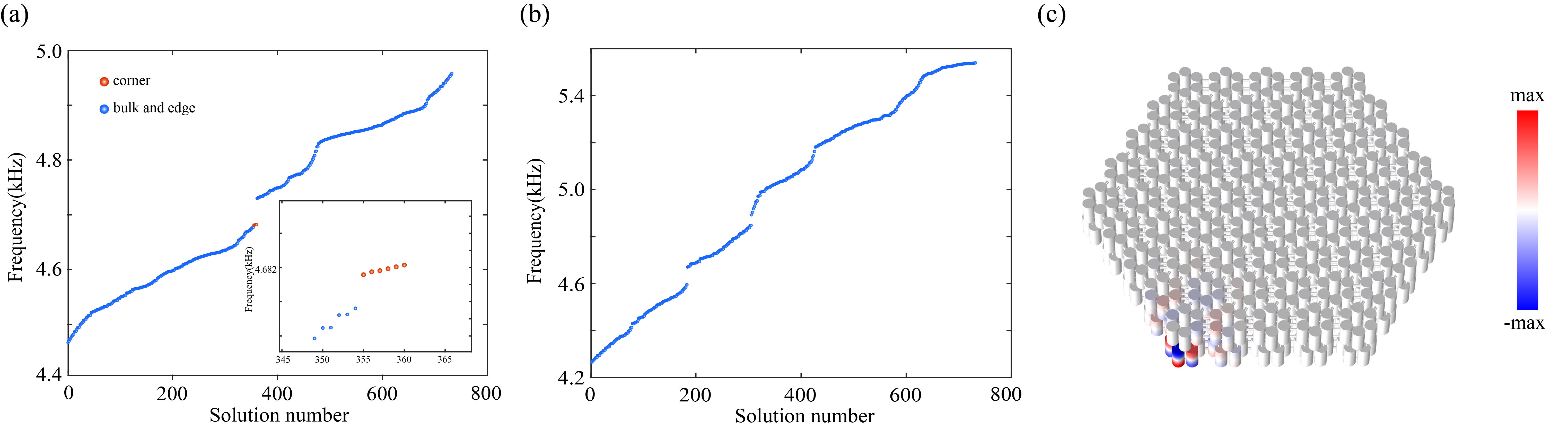}
		\caption{Eigen spectra of the $C_6$-symmetric HBPCs with (a) weak interlayer coupling $d_3=1.2$ mm, and (b) strong interlayer coupling $d_3=8.0$ mm. (c) Simulated acoustic pressure profile the corner state marked in red points in (a).}
		\label{Fig_14}
	\end{figure*}

	It is known that a two-dimensional higher-order topological phases promise a symmetry-protected zero-dimensional corner states. To confirm this, we construct a finite-sized HBPC [see schematically shown in Fig.~\ref{Fig_13}(a)] and calculate its eigen spetrum with both weak and strong interlayer coupling. For HBPCs with weak interlayer coupling, i.e., $d_3=1.2$ mm, as expected, six corner states marked by red points are identified at $f=4682$ Hz in the nontrivial band gap in Fig.~\ref{Fig_14}(a). The corresponding acoustic pressure field distributions are displayed in Fig.~\ref{Fig_14}(c). Interestingly, since most of the field are localized in the lower layer configured with $h^{(6)}_{3c}$ type, it is believed that the nontrivial topology mostly originated from the monolayer with $h^{(6)}_{3c}$ configuration. In contrast, for HBPCs with strong interlayer coupling, namely,$d_3=8.0$mm, there is no corner state emerge in the first bandgap owing to the trivial band topology [see Fig.~\ref{Fig_14}(b)].
	
	\subsection{$C_3$-symmetric HBPC formed by upward and downward lattices ($h^{(3)}_{2b}$ and $h^{(3)}_{2c}$)}
	
\begin{figure*}[htbp]
		\centering\includegraphics[width=6.8in]{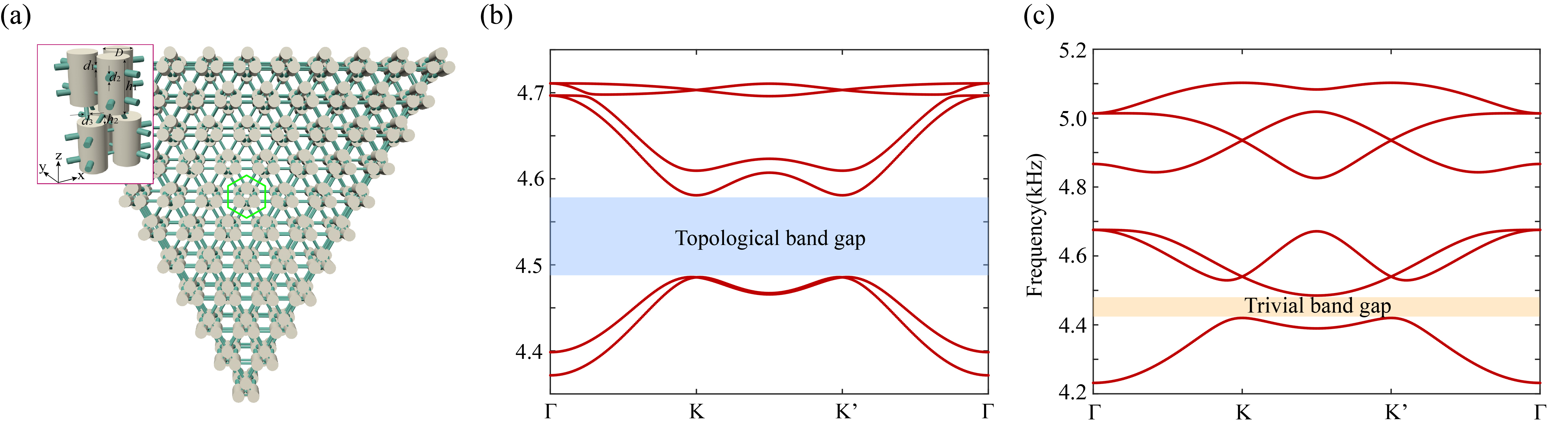}
		\caption{ (a) Top view of the $C_3$-symmetric HBPC, which formed by connecting the monolayers phononic crystal arrayed with Wu-Hu's and hexagon lattices ($h^{(3)}_{2b}$ and $h^{(3)}_{2c}$), respectively. Inset: the unit cell. The geometric parameters of two monolayer phononic crystals are adopted from Figs.~\ref{Fig_8}(a) and ~\ref{Fig_11}(a), respectively. (b,c) Simulated band structure of HBPC with (b) weak interlayer coupling, i.e., the diameter of the connecting air tube $d_3=1.2$ mm, and (c) strong interlayer coupling, i.e., the diameter of the connecting air tube $d_3=8.0$ mm.}
		\label{Fig_15}
	\end{figure*}

	Following the above procedure, we then consider a $C_3$-symmetric HBPC formed by stacking upward and downward lattices with nontrivial band topology. To this end, we adopt the monolayer phononic crystals configured with $h^{(3)}_{2b}$ and $h^{(3)}_{2c}$, which the geometric parameters are same with the Figs.~\ref{Fig_7}(a) and ~\ref{Fig_10}(a), respectively. As shown in Fig.~\ref{Fig_15}(a), the unit cell indicated by the green hexagon consisting of six acoustic cavities, each cavity is connected with two nearest neighbour cavities (within the unit cell) of upper or lower layer via two sloped air tubes. We then study the effect of interlayer coupling on the band topology in HBPCs with different interlayer coupling. As shown in Figs.~\ref{Fig_15}(b) and ~\ref{Fig_15}(c), we present the band structure of the HBPC with $d_3=1.2$mm and $d_3=8$mm, which represent the weak and strong interlayer coupling, respectively. Under the weak interlayer coupling, namely $d_3=1.2$mm, it is seen that a band gap divide the twelve bands into two sets, the lower bandset below the gap consisting of two bands, while higher bandset above the gap consisting of four bands. Interestingly, the band structure of the $C_3$-symmetric HBPC can be regard as the superposition of the band structure of the monolayer configured with $h^{(3)}_{2b}$ and $h^{(3)}_{2c}$ types. Nevertheless, due to the absence of the mirror symmetry, the separability into subspaces with opposite parities is no longer valid in HBPCs, which is in strong contrast to that in the $C_3$-symmetric MBPC. 	
	
	On the other hand, the band gap of the monolayer phononic crystal with $h^{(3)}_{2c}$ and $h^{(3)}_{2b}$ configuration share the same frequency range, which, however, are of different nontrivial topology. Following Ref.~\cite{topocharater1}, the topological crystalline index of $C_3$-symmetric MBPC can be expressed by the full set of the $C_3$ eigenvalues at the high-symmetry points (HSPs). For an HSP denoted by the symbol $\Pi$, the $C_3$ eigenvalues can only be $\Pi_n=e^{i2\pi(n-1)/3}$ with $n$ range from 1 to 3. Here, the HSPs include the $\Gamma$, $K$ and $K^\prime$. The minimum set of indices that describe the band topology of $C_3$ crystalline insulator is given by 
	\begin{equation}
		\left[ K^{(3)}_n \right] = \# K^{(3)}_n -\# \Gamma^{(3)}_1, n=1,2,3,
	\end{equation}
	where $\# K^{(3)}_n (\# \Gamma^{(2)}_n)$ refers to the number of bands below the band gap with the $C_3$ symmetry eigenvalue $K_n (\Gamma_n)$ at the $K (\Gamma)$ point. Hence, the property of the gap can be characterized by 
	\begin{equation}
		\xi^{(3)}=\left ( \left [ K_1^{(3)}\right ], \left [K^{(3)}_2 \right ] \right ).
	\end{equation}
	The calculation show the topological index of the band gap is $(2,-1)$, indicating a nontrivial higher-order topology. Hence, it is emphasized that the the band topology of the HBPC inherits from the nontrivial monolayer phononic crystal under weak interlayer coupling. 
	
	We proceed to discuss the band structure of the HBPC under the strong interlayer coupling, namely $d_3=8$mm, in Fig.~\ref{Fig_15}(c). It is seen that two band gap divide the twelve bands into three bandsets. To be specific, there is only a single band below the first band, while higher bandsets are reorganized owing to the strong interlayer coupling, which makes a difference from that weak interlayer coupling. Here we only focus on the first bandgap indicated by the orange area in Fig.~\ref{Fig_15}(c). Remarkably, the calculated topological crystalline index gives $\xi^{(3)}=(0,0)$, indicating a trivial phase. Hence, it is emphasized that the phase transition in the HBPC can be triggered by adjusting the interlayer couplings, rather than breaking the geometric symmetry.
	
\begin{figure*}[htbp]
		\centering\includegraphics[width=6.8in]{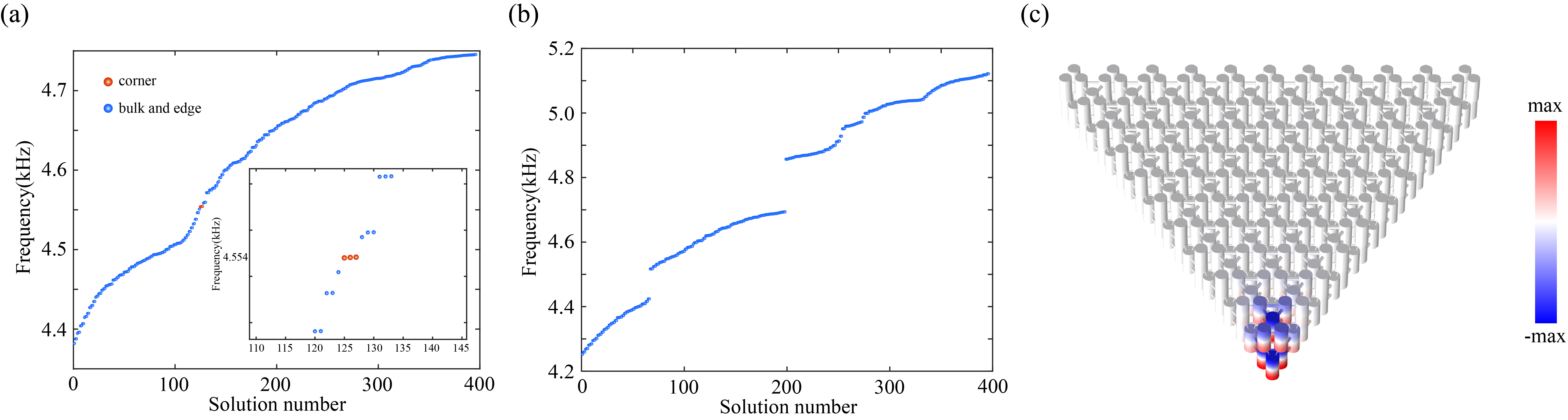}
		\caption{Eigen spectra of the $C_3$-symmetric HBPCs with (a) weak interlayer coupling $d_3=1.2$ mm, and (b) strong interlayer coupling $d_3=8.0$ mm. (c) Simulated acoustic pressure profile the corner state marked in red points in (a).}
		\label{Fig_16}
	\end{figure*}

	It is known that a two-dimensional higher-order topological phases promise a symmetry-protected zero-dimensional corner states. To confirm this, we construct a finite-sized HBPC [see schematically shown in Fig.~\ref{Fig_15}(a)] and calculate its eigen spetrum with both weak and strong interlayer coupling. For HBPCs with weak interlayer coupling, i.e., $d_3=1.2$ mm, as expected, three corner states marked by red points are identified at $f=4554$ Hz in the nontrivial band gap in Fig.~\ref{Fig_16}(a). The corresponding acoustic pressure field distributions are displayed in Fig.~\ref{Fig_16}(c). In contrast, for HBPCs with strong interlayer coupling, namely,$d_3=8.0$mm, there is no corner state emerge in the first bandgap owing to the trivial band topology [see Fig.~\ref{Fig_16}(b)].
	
	\section{Conclusion and discussions}
	In conclusion, we systematically studied the $C_6$- and $C_3$-symmetric higher-order topological phases in bilayer phononic crystals with two types of stackings: the mirror symmetric stacking and the heterogeneous stacking. For the mirror-stacked bilayer lattice, the separability of the Hilbert space with odd and even parities and the tuning of the interlayer coupling enables the emergence and disappearance of the topological corner states in the bulk continuum. For the bilayer phononic crystals formed by two distinct monolayer lattices of the same symmetry, the band topology is strongly affected by the interlayer couplings as well as the two original monolayers. The bilayer phononic crystals can even experience a phase transition from nontrivial to trivial band topology when the interlayer coupling is gradually strengthened. Our work unveil the rich physics of the interlayer couplings in bilayer systems, suggesting that layer degree of freedom can be used to enrich the higher-order topological phases.
	
	\section*{ACKNOWLEDGMENTS}
	This work was supported by the National Key R\&D Program of China (2022YFA1404400), the National Natural Science Foundation of China (Grant Nos. 12125504, 11904060, 12074281, and 12204417), the ``Hundred Talents Program'' of the Chinese Academy of Sciences, the Natural Science Foundation of Guangxi Province (Grant No. 2023GXNSFAA026048), the Priority Academic Program Development (PAPD) of Jiangsu Higher Education Institutions, and the project of all-English course construction for graduate students in Guangxi Normal University.
	
	\bibliography{Bilayer}
	
\end{document}